\documentclass{sig-alternate-05-2015}

\usepackage[ruled,vlined,linesnumbered]{algorithm2e}
\usepackage{amsmath}

\usepackage{times}
\usepackage{helvet}
\usepackage{courier}
\usepackage{graphicx}
\usepackage{balance}
\usepackage{subcaption}
\usepackage{booktabs}
\usepackage{multirow}

\usepackage{caption}

\begin{document}

\CopyrightYear{2016} 
\setcopyright{acmcopyright}
\conferenceinfo{CIKM'16 ,}{October 24-28, 2016, Indianapolis, IN, USA}
\isbn{978-1-4503-4073-1/16/10}\acmPrice{\$15.00}
\doi{http://dx.doi.org/10.1145/2983323.2983803}

% Copyright
\setcopyright{acmcopyright}
%\setcopyright{acmlicensed}
%\setcopyright{rightsretained}
%\setcopyright{usgov}
%\setcopyright{usgovmixed}
%\setcopyright{cagov}
%\setcopyright{cagovmixed}

% DOI
%\doi{10.475/123_4}

% ISBN
%\isbn{123-4567-24-567/08/06}

%Conference
%\conferenceinfo{PLDI '13}{June 16--19, 2013, Seattle, WA, USA}

%\acmPrice{\$15.00}

%
% --- Author Metadata here ---
%\conferenceinfo{WOODSTOCK}{'97 El Paso, Texas USA}

\title{Mining Shopping Patterns for Divergent Urban Regions by Incorporating Mobility Data\thanks{This work was conducted while the first author was doing internship at
Microsoft Research Asia}}

\vspace{-0.1in}

\numberofauthors{1}
\author{%
	  \alignauthor Tianran Hu$^{*1}$, Ruihua Song$^{\dag2}$, Yingzi Wang$^{{\dag\mathcal{x}}3}$, Xing Xie$^{\dag4}$, Jiebo Luo$^{*5}$\\
	  \affaddr{
	  	$^*$University of Rochester
	  	$^\dag$Microsoft Research\\
	  	$^{\mathcal{x}}$University of Science and Technology of China  	
	  }\\
	  \email{
	  \{$^1$thu, $^5$jluo\}@cs.rochester.edu,\\ \{$^2$song.ruihua, $^4$xing.xie\}@microsoft.com,	$^3$yingzi@mail.ustc.edu.cn
	  }
}

\maketitle
\begin{abstract}
What people buy is an important aspect or view of lifestyles. 
%Shopping pattern is an important view of human lifestyles. 
Studying people's shopping patterns in different urban regions can not only provide valuable information for various commercial opportunities, but also enable a better understanding about urban infrastructure and urban lifestyle. In this paper, we aim to predict city-wide shopping patterns. This is a challenging task due to the sparsity of the available data -- over 60\% of the city regions are unknown for their shopping records. To address this problem, we incorporate another important view of human lifestyles, namely mobility patterns. With information on ``where people go'', we infer ``what people buy''. Moreover, to model the relations between regions, we exploit spatial interactions in our method. To that end, Collective Matrix Factorization (CMF) with an interaction regularization model is applied to fuse the data from multiple views or sources. Our experimental results have shown that our model outperforms the baseline methods on two standard metrics. Our prediction results on multiple shopping patterns reveal the divergent demands in different urban regions, and thus reflect key functional characteristics of a city. Furthermore, we are able to extract the connection between the two views of lifestyles, and achieve a better or novel understanding of urban lifestyles.

\end{abstract}


\section{Introduction}
%Regions in an urban area usually have divergent functions~\cite{}. 
Studies on land use suggest a variety of types of city regions, such as residential, industrial and others~\cite{xiao2006evaluating,antikainen2005concept}. Such divergence of city region functionality ~\cite{yuan2012discovering} naturally leads to the various shopping patterns in different regions. Daily supplies, for example, are usually needed in residential quarters, while office appliances are more popular in business areas. In this paper, we aim to predict shopping patterns in different urban regions. Through the prediction, we are able to read the various demands of these regions, gain knowledge of the infrastructures of urban areas, and more importantly, acquire a better understanding of human lifestyle geographically. Furthermore, this study could be used to support location-based marketing, such as mobile advertising, outdoor exhibition stands, and so on.

Market Basket Analysis %han2011data
shows that consumers usually have demands for a group of products~\cite{brin1997dynamic}, and people's demands are highly related to their lives~\cite{du2007profiling}. 
%Consumer Lifestyles in the US
%shopping patterns are highly related to their lives. 
For example, an IT practitioner may have interests in digital products, computer games, as well as technical books. Motivated by this observation, we consider shopping patterns as combinations of inherently related types of products. Our goal in this study is to predict the levels of shopping patterns in different city regions. Compared with predicting the levels of single types of products, there are two main benefits to doing so. First, since products are actually correlated, the prediction of a group of related products provides more valuable information for item recommendation and commercial planning. Second, since shopping patterns are more related to the lifestyles of people in a region than single categories of products, predicting shopping patterns can give us a more comprehensive understanding of urban areas. 

In this study, we collect geolocated logs of browsing products online from Beijing, and extract shopping patterns for the regions in the city. Although the visiting volumes to product webpages are not the exact product trading volumes, they indicate the demands for different products. Therefore, the browsing records can be regarded as reasonable proxy of consumer shopping behaviors. According to the study on human mobility~\cite{ye2011exploiting}, the probability of a person moving a certain distance drops dramatically as the distance exceeds 1km in urban area. Therefore, we define a region as a 1km by 1km grid along the latitude and longitude lines in a city. 
%Regions' shopping patterns are represented by a region by shopping pattern matrix, and the elements in the matrix denotes the level of a shopping pattern in a region.
We formalize region-based shopping patterns as a matrix, and an element in the matrix denotes the level of a shopping pattern in a region. Our task is to complete the missing values in the matrix. The matrix suffers from a severe sparsity problem -- over 60\% of rows are empty. In other words, we do not have any shopping information of more than 60\% regions in the city. To address this problem, we seek help from other types of lifestyle information.

\begin{figure}[!htbp]
\centering
\includegraphics[width=1 \columnwidth]{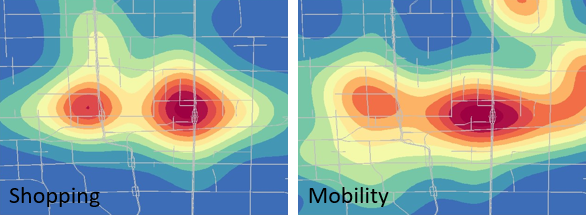}
\caption{The illustration of the geographical distributions of a shopping pattern (left) and a mobility pattern (right) of the same region on a local map in Beijing.}~\label{fig:compare}
\vspace{-2em}
\end{figure}

\textit{Lifestyle} is defined as ``\textit{the way in which a person or a group of people live}'' in Merriam-Webster Dictionary\footnote{http://www.merriam-webster.com/dictionary/lifestyle/}. There are many aspects or views to describe how people live~\cite{wang2015your,yuan2013we}, including shopping patterns and mobility patterns. For example, a teenage student may have a certain shopping pattern, such as books, digital products, and so on. Similar to shopping patterns, we consider a mobility pattern as a group of inherently related POI categories. We can also see a certain mobility pattern of the student, such as dormitory, library, and so on. Multiple views of a lifestyle are actually connected, and this connection is reflected on their geographical distributions. Figure~\ref{fig:compare} shows the geographical distributions of a shopping pattern extracted from the browsing logs and a mobility pattern extracted from check-in data in a region. The top three most browsed product categories of the illustrated shopping pattern (left) are Laptops, Office Appliances \& Stationery, and Tablets. The top three most checked in POI categories of the mobility pattern (right) are Offices, Convenience Stores, and Commute. We observe that the geographical distributions of these two patterns are quite similar, and they both reach the highest levels in the same block. In other words, many people check in at offices (mobility) in this block, meanwhile, many people show high demands for office appliances (shopping) in the same block. The geographical connection of these two views of the lifestyle inspires this work: connecting shopping and mobility patterns, and utilizing the mobility patterns in a region to predict the shopping patterns there. To achieve this, we learn region mobility patterns from check-in data of a popular Location Based Social Network (LNSB) in China, and employ a Collective Matrix Factorization (CMF)~\cite{singh2008relational} model to fuse these two views.

However, considering the lifestyles of regions in a city separately is inappropriate. Regions in a city are never isolated, as they are connected by people transitions instead~\cite{chawla2012inferring,zheng2015detecting}. Intuitively, the commuting flows between regions also imply the flows of lifestyles. For example, it is reasonable to assume that consumers' demands affect the shopping patterns in both the regions where they live and work. We assume that the stronger the interactions between two regions are, the more alike the lifestyles in both regions are. We define the interaction from a region to another region as the probability of a person moving from the former to the latter. We first collect two large human transition datasets in Beijing, and then apply a gravity model~\cite{wilson1967statistical} to quantify the interactions. We incorporate the interactions between regions as an regularization term in our overall model. In comparison with several baseline methods, our model improves the accuracy by 10.8\% and 12.6\% in terms of the Mean Absolute Error (MAE) with 80\% and 90\% training data, respectively.

Our main contributions are summarized as follows:
\begin{itemize}

\item \textit{Dealing with data sparsity}: %The shopping pattern data is very sparse, over 60\% regions are unknown. To address this issue, 
We incorporate mobility patterns of regions as complementary information to address the sparsity issue of our shopping data. We leverage the underlying connection between mobility and shopping patterns, and use ``where people go'' to help infer ``what people buy''. Moreover, to model the relations between regions and further make up for the spare data, we consider the interactions between regions in our overall model. A hybrid model is proposed to utilize several data sources or views. The experimental results validate that our model outperforms the baseline methods according to two standard metrics. 

\item \textit{Multi-view Lifestyles}: We study the connection between different views of lifestyle particularly in terms of shopping and mobility patterns. The reasonable mappings between these views guarantee the effectiveness of incorporating mobility patterns to predict the shopping patterns. Moreover, interesting mappings between two views provide a better and novel understanding of human lifestyles in the city.

\item \textit{City-wide shopping patterns modeling}: We extract representative shopping patterns of regions in urban areas, and predict the level of each shopping pattern. Our results show the dramatic divergence in shopping demands over the city. This work sheds light on a better understanding of the urban infrastructure, and provides valuable information for various business opportunities. 

\end{itemize}

\section{Related Work}
\subsection{Urban Region Divergences}
Previous works on urban land use suggest various types of city regions~\cite{xiao2006evaluating,antikainen2005concept}. Singh et. al report their findings on noise levels of three types of city regions: residential, industrial, and commercial~\cite{singh1995comparative}. As social media has been being widely adopted in people's daily lives, social media data becomes an important data source for related research. For example, geolocated tweets are used in~\cite{frias2012characterizing} to characterize urban landscapes. In this work, the authors use tweet contents to identify business, leisure, nightlife, and residential regions in New York City. 

Since urban data is usually noisy and sparse, researchers have to combine several data sources to dig meaningful knowledge. Yuan et. al~\cite{yuan2012discovering} discover functional zones in Beijing. In this work, the authors cluster regions into several functional groups based on their transition patterns. POI information in a region is used as meta data for the region in this work. In~\cite{zheng2014diagnosing}, Yu et. al report their work on predicting the noise levels of New York City. City road structures, POI distribution, and check-in data are used in this work. Interesting correlations are found between multiple data sources and noise level. For example, universities in a region are quite responsible for ``party \& loud speak'' noise. Similarly, the authors of \cite{shang2014inferring} introduce road network and POI locations to infer the traffic volume in a region.
%Similarly, \cite{shang2014inferring} and \cite{zheng2013u} utilize several data sources to infer pollution emission and air quality in urban area, respectively. 
%The authors of the former one introduce road network and POI locations to infer the traffic volume in a region. In the later one, the authors find that air quality in a region is not only the result of meteorological factors, but also affected by human activities there.
\cite{zheng2015methodologies} gives a nice survey on cross-domain data fusion in urban computing.

\subsection{Human Lifestyles}
Human lifestyles are well studied in sociology~\cite{ansbacher1967life,kamphorst1990leisure}. In recent years, researchers have successfully utilized social media in research ventures related to lifestyle analysis~\cite{sadilek2013modeling,abbar2014you}. For example, Noulas et al. of~\cite{noulas2011empirical} use Foursquare data to discover the behavioral habits of residents in London. Based on tweet contents, Sadiek et al. build a language model to detect the health of individuals~\cite{sadilek2013nemesis}. 

Much effort has been made to linking multiple types of lifestyles. Cranshaw et. al~\cite{cranshaw2010bridging} report that human mobility patterns are strongly connected to their social networks. In~\cite{zhong2015you}, the authors use check-in data to infer users' demographic features. Furthermore, Yuan et. al linked several aspects of human lives, such as occupation, education level, and mobility patterns by studying the data from several social media platforms~\cite{yuan2013we}. Inspired by these works, our key idea in this work is using information of one view (mobility) of lifestyles of a region to infer the other one (shopping).

\subsection{Matrix Factorization Techniques}
Matrix Factorization (MF) techniques, also known as Matrix Decomposition, are widely applied in recommendation systems to predict missing values in matrices~\cite{sarwar2001item}. A basic MF model factorizes the target matrix into a latent pattern matrix and a coefficient matrix, and uses the product of them as the estimation to missing values. 

However, this approach usually suffers from data sparsity problems~\cite{koren2009matrix} -- the information contained in the original matrix is not enough for a precise factorization. Our dataset has such a problem -- a majority of rows in our target matrix are empty. Many techniques have been developed to deal with this problem~\cite{park2009pairwise,zhou2011functional}, including Collective Matrix Factorization (CMF)~\cite{singh2008relational}. In a CMF model, supplementary information is provided in terms of matrices. Instead of decomposing only the target matrix, several matrices are decomposed together. By doing so, supplementary information helps to modify the factorization results, and produce a better prediction result. In this work, we apply a CMF model to combine shopping and mobility patterns. Social regularization is reported in~\cite{ma2011recommender}. In this work, Ma et. al consider the social connections between users in a recommendation system, and introduce the connections into a transitional MF model using a social regularization term. We introduce region interactions into our model using an interaction regularization term. The interactions are calculated using a gravity model~\cite{wilson1967statistical}. This model assumes the human mobility in a city follows Newton's law of gravity, and is widely applied in urban geography. Wang et. al~\cite{wang2015regularity} employ this model to calculate the influence between regions in their POI recommendation system. 

\section{Overview}

\begin{figure}[!htbp]
\centering
\includegraphics[width=1\columnwidth]{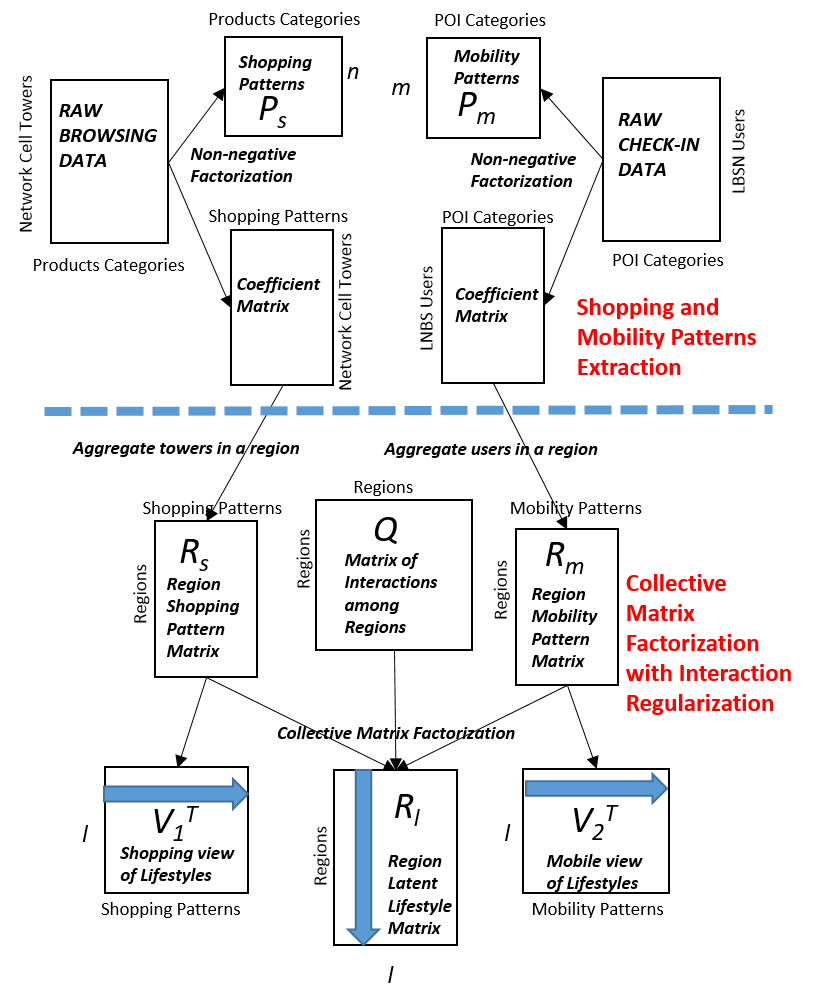}
\caption{Our framework}~\label{fig:workflow}
\vspace{-2em}
\end{figure}

Before we introducing our model, we first go through the workflow of this study in Figure~\ref{fig:workflow}. There are two main parts to this work: 1) shopping and mobility patterns extraction, and 2) predicting shopping patterns using a collective matrix factorization model with interaction regularization. 

The first part starts from the raw browsing data and raw check-in data. We extract $n$ shopping patterns (matrix $P_s$) and $m$ mobility patterns (matrix $P_m$) from these two datasets. $n$ and $m$ are input parameters. From this process, we also obtain the weights on shopping patterns of different locations, as well as the weights on mobility patterns of LBSN users in the form of two coefficient matrices.

The second part starts from aggregating the locations and users into regions. After the aggregation, we obtain a region shopping pattern matrix $R_s$, a $r$ by $n$ matrix, and a mobility pattern matrix $R_m$, a $r$ by $m$ matrix. Here, $r$ denotes the number of regions in the city. Also, we prepare the interaction matrix $Q$, a $r$ by $r$ matrix, which contains the interaction information between regions. We then apply collective matrix factorization with interaction regularization on $R_s$, $R_m$, and $Q$. This is based on the assumption that $P_s$ can be approximately factorized to $R_l \times V_1^{\top}$, meanwhile, $P_m$ to $R_l \times V_2^{\top}$. Three matrices are generated from this process: 
%$R_l$, region latent lifestyle matrix, $V_1$, shopping view of latent lifestyles, and $V_2$, mobility view of each latent lifestyles.
1) $R_l$, a $r$ by $l$ region latent lifestyle matrix, which indicates the weights on latent lifestyles of regions. $l$ is an input parameter denoting the number of latent lifestyles. 2) $V_1^{\top}$, a $l$ by $n$ matrix. This matrix is the shopping view of latent lifestyles. 3) $V_2^{\top}$, a $l$ by $m$ matrix, contains the mobility view of latent lifestyles. We predict the missing values in $R_s$ with the corresponding elements of the product of $R_l$ and $V_1^{\top}$.

The basic idea behind collective factorization here is assuming that the weights on shopping patterns $R_s$ and mobility patterns $R_m$ of regions are both generated by $l$ latent lifestyles. Each row in $V_1^{\top}$ explains a latent lifestyle with weights on shopping patterns, while a row in $V_2^{\top}$ explains a latent lifestyle with weights on mobility patterns. This process also reveals exquisite mappings between human shopping behaviors and movements. There are $l$ columns in $R_l$ corresponding to $l$ latent lifestyles. The $i$th column of $R_l$, for example, denotes the weights of regions on the $i$th row of $V_1^{\top}$ when using these two matrices to recover $R_s$. Meanwhile, $i$th column of $R_l$ also denotes the weights of the regions on the $i$th row of $V_2^{\top}$ when using these two matrices to recover $R_m$. This intuitively implies the mapping between $i$th row in $V_1^{\top}$ and $V_2^{\top}$ -- they are the two views of the same latent lifestyle. 

\section{Shopping and Mobility patterns of Regions}

We define a shopping pattern as a group of related product categories. For example, Furniture, Home Textiles, Large Home Appliances, and so on are extracted as a shopping pattern in this work. A mobility pattern is a group of related POI categories, such as a combination of School Dormitories, School Libraries, and so on. Shopping and mobility patterns are basic bricks of two views of lifestyles. In this section we explain how we extract these patterns.

\subsection{Shopping Patterns Extraction}
We collect the browsing log of shopping websites from the city of Beijing. Each log record is associated with a product category and a location identifier. There are no user identifiers in the data. Moreover, the location identifiers are not the precise locations where consumers visit product webpages, but the locations of the network cell towers their devices connect to. 

Since we cannot identify users in our data, as a compromise, we use the data from a network cell tower as the basic unit for shopping pattern mining. A a cell tower covers a relatively small area, hence, we assume that the people connected to a tower have similar shopping patterns. Note that, a city region could contain multiple or no network cell towers. We first aggregate the browsing records from the same locations into a vector, we name this vector the browsing vector of a location. The length of a browsing vector is the number of product categories, and the value of a certain element is the browsing amount to that product category from a location. The browsing vectors are accumulated to form a browsing matrix. A row in this matrix indicates a location (a network cell tower), and a column indicates a product category. 
%The amount of rows equals to the number of cell towers in the city. 
We then apply a Nonnegative Matrix Factorization (NMF) model on this browsing matrix, since NMF leads to interpretable results~\cite{lee1999learning}. We obtain two result matrices through the processes: a shopping pattern matrix $P_{s} \in \mathbb{R}^{n \times c_s}$ and a coefficient matrix, where $n$ is the number of shopping patterns, and $c_s$ is the number of product categories. A row of $P_{s}$ indicates the weights on product categories of a shopping pattern. The coefficient matrix indicates the weights on each shopping pattern of locations. 

\begin{figure}[!htbp]
\vspace{-1em}
\centering
\includegraphics[width=1\columnwidth]{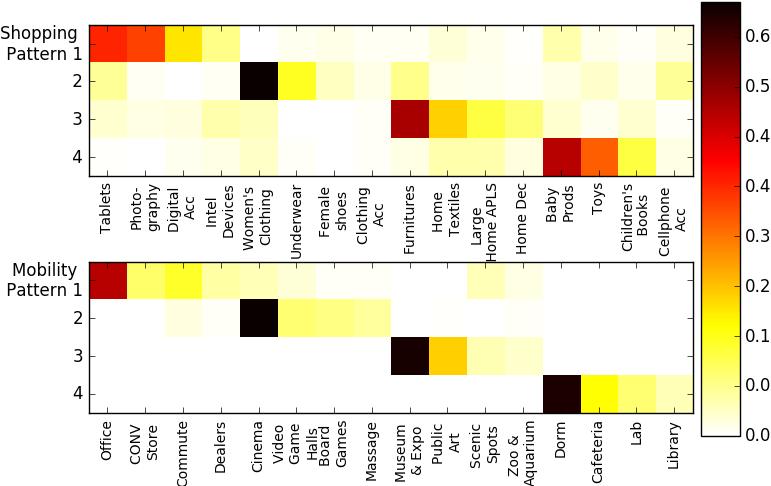}
\caption{The top weighed categories of four shopping patterns (upper), and four mobility patterns (lower). Horizontal axises indicate product categories and POI categories, respectively. ``Accs'', ``Conv'', ``Dorm'', ``APLS'', ``Intel'', ``Prods'', and ``Dec'', are short for ``Accessories'', ``Covenient'', ``Dormiatroies'', ``Applicances'', ``Intellegent'', ``Products'', and ``Decoration'', respectively.}~\label{fig:life_patterns}
\vspace{-2em}
\end{figure}

In Figure~\ref{fig:life_patterns} we plot the top weighted product categories of several shopping patterns (upper part). Please note that the amount of categories of products and POIs is more than what we display in the figure. We can see that each shopping pattern has a specific emphasis on a group of related products, which naturally represents a class of demands of people. The first shopping pattern mainly focuses on digital products. Tablets, Photography Products, and Digital Accessories are the top three most weighted categories. 
%The second presents a high demand for female fashion. 
The second pattern is dominated by Women's Clothing, and other categories such as Underwear and Female Shoes have certain shares in the pattern. The third pattern shown in the figure is mainly about home arrangements and decorations, Furniture, Home Textiles, and Large Home Appliances are the most popular product categories in this pattern. The last shopping pattern has highest weights on Baby Products, Toys, and Children's Books. 

We then sum up the weights on shopping patterns of all the locations in a region to obtain the weights on the shopping patterns of a region. The result is represented by a matrix $R_{s} \in \mathbb{R} ^ {r \times n}$, where $r$ is the number of regions in the city. The rows of this matrix indicate regions, and columns represent shopping patterns. Since many regions do not contain records at all, $R_{s}$ is a sparse matrix. Our goal in this work is to fill in the empty elements of $R_s$.

\subsection{Mobility Patterns Extraction}
To extract human mobility patterns, we collect check-in data form Jiepang\footnote{http://www.jiepang.com/}, a popular LBNS in China. Each check-in is associated with a user ID, a POI category, and a latitude and longitude pair. We take a user as the basic unit for mobility pattern mining. The process is similar to shopping pattern extraction. We first aggregate users' check-ins into vectors, named mobility vectors. Mobility vectors record users' amounts of check-ins of all POI categories. We then accumulate mobility vectors into a mobility matrix, and apply NMF on it to extract a mobility pattern matrix $P_{m}  \in \mathbb{R}^{m \times c_m}$, and a coefficient matrix. Here, $m$ denotes the amount of mobility patterns, and $c_m$ is the number of POI categories. $P_{m}$ describes the ways in which people live from a view of spatial activities, and the coefficient matrix records the weights of users on these mobility patterns.

Figure~\ref{fig:life_patterns} (lower part) shows four mobility patterns with their most weighted POI categories. Similar to shopping patterns, a mobility pattern also focuses on a specific group of categories. 
%while the difference is that the most weighted category in a mobility pattern appears more dominant. 
The first pattern indicates a mobility pattern centered at Offices, followed by Convenient Stores and Commute. 
%The high weighted POI categories are quite related to people's daily working. 
The second mobility style is dominated by Cinemas. The next two categories are Video Game Halls and Board Games. This is clearly an entertainment related mobility pattern. The third one is more about outdoor activities, which assigns higher weights to Museum \& Exhibition. The last mobility pattern mainly focuses on POIs related to schools or universities. Dormitories, Laboratories, and Libraries have high weights in this pattern.

Since LBNS users may leave footprints in several regions, it makes more sense to allocate the same proportion of user mobility patterns to the regions they checked in with the percentages of activities they spent in these regions. For example, if a user spends 40\% check-ins in a region, then this user contributes 40\% of her mobility patterns to this region. Therefore, we define the mobility patterns of a region as the weighted sum of all the users' mobility patterns, who checked in this regions before. Matrix $R_m \in \mathbb{R}^{r \times m} $ is then formed, where the element on $i$th row and $j$th column denotes $i$th region's weight on $j$th mobility pattern. Formally,

\begin{equation}
R_{m_{i,j}} = \displaystyle\sum_{k}^{} w_{k,i} \times u_{k,j},
\end{equation} where $w_{k,i}$ is the percentage of activity which user $k$ spends in region $i$, and $u_{k,j}$ is this user's weight on $j$th mobility pattern.

\section{Hybrid Model for Shopping Pattern Prediction}
$R_{s}$ contains the shopping pattern information of regions, and is a sparse matrix. Our objective is to fill the empties so as to predict shopping patterns of missing regions. In this section, we apply a Collective Matrix Factorization model with an interaction regularization term to achieve the goal.

\subsection{Collective Matrix Factorization}
%$R_{s}$ and $R_{m}$ contain the shopping and mobility information of regions. 
We treat shopping and mobility patterns in a region as two different views generated by same lifestyles. In other words, we assume $R_s$ and $R_m$ are both generated from a matrix containing the latent lifestyle information of each region. Formally,

\begin{align}\label{equ:approx}
R_{s} \approx R_{l} V_{1}^\top\\
R_{m} \approx R_{l} V_{2}^\top
\end{align} where $R_{l} \in \mathbb{R} ^ {r \times l}$ denotes the weights on $l$ latent lifestyles of each region. $V_{1}^{\top} \in \mathbb{R} ^ {l \times n}$ explains $l$ latent lifestyles with $n$ shopping patterns. Similarly, $V_{2}^{\top} \in \mathbb{R} ^ {l \times m}$ explains $l$ latent lifestyles with $m$ shopping patterns. Guided by the extra mobility information of $R_{m}$, we obtain more accurate $R_{l}$ and $V_{1}$. The prediction of missing values in $R_{s}$ is achieved by multiplying $R_{l}$ and $V_{1}$. Following this idea, we apply a collective matrix factorization model to decompose $R_{s}$ and $R_{m}$ together. The objective function is formulated as:

\begin{align}
L(R_{l}, V_{1}, V_{2}) = \frac{1}{2} \Vert  I \circ (R_{s} - R_{l}V_{1}^\top) \Vert_{F}^2 + \frac{\lambda_{1}}{2} \Vert R_{m} - R_{l}V_{2}^\top \Vert_{F}^2\notag
\\+ \frac{\lambda_{2}}{2}(  \Vert R_{l} \Vert_{F}^2 +  \Vert V_{1} \Vert_{F}^2 +  \Vert V_{2}  \Vert_{F}^2) ,
\end{align} where operator ``$\circ$'' means the entry-wise product, $\Vert.\Vert_{F}^2$ denotes the Frobenius norm, and $I$ is an indicator matrix with its entry $I_{ij} = 0$ if $R_{s_{ij}}$ is missing. The last form is for preventing overfitting. $\lambda_1$ and $\lambda_2$ are parmeters controlling the contribution of each part. 

An interesting knowledge we can learn from this approach is the mapping between two views. The $i$th row of $V_{1}^\top$ and $V_{2}^\top$ are both associated with the $i$th column in $R_{l}$, which implies that they represent the same latent lifestyle. However, only associating two views of lifestyles ignores the interactions between regions. We quantify the interactions using a gravity model, and introduce this factor in the objective function as a regularization term.

\subsection{City-wide Interaction Regularization}

\subsubsection{Gravity Model}
We define the interaction from region $i$ to region $j$ as the probability of a person moving from $i$ to $j$. To compute this probability, we employ and adapt a gravity model~\cite{wilson1967statistical}, which is widely adopted in mobility analytics for a large population. 

The gravity model states that the commuting flows from region $i$ to $j$, denoted as $q_{i,j}$, are determined by 1) $O_i$, the number of individuals leaving region $i$, 2) $D_j$, the number of individuals arriving at region $j$, and 3) the distance between two regions, through a gravity-like law:

\begin{equation}\label{equ:gravity_model}
q_{i,j} = c \frac{(O_{i})^{a}(D_{j})^{b}}{\exp(g \cdot dis_{i,j})},
\end{equation} where $c$ is a constant, $a$ and $b$ are coefficients of $O_i$ and $D_j$, respectively. $g$ is a parameter that tunes the decay by distance. When $i = j$, this model computes the interactions within a region.

%There are also intra-region interactions indicating the movements take place in single regions. This type of interactions is also modeled with this model when $i = j$. 

We fit this model using observed transition data to estimate the coefficients $a$, $b$, and $g$. Applying a logarithmic transformation to both side of (\ref{equ:gravity_model}), we obtain the following expression:

\begin{equation}\label{equ:estimation1}
\ln{q_{i,j}} = a\ln{{O}_{i}} + b\ln{{D}_{j}} - g\cdot dis_{i,j} + \ln{c}
\end{equation}

\begin{figure}[!htbp]
\centering
\includegraphics[width=1 \columnwidth]{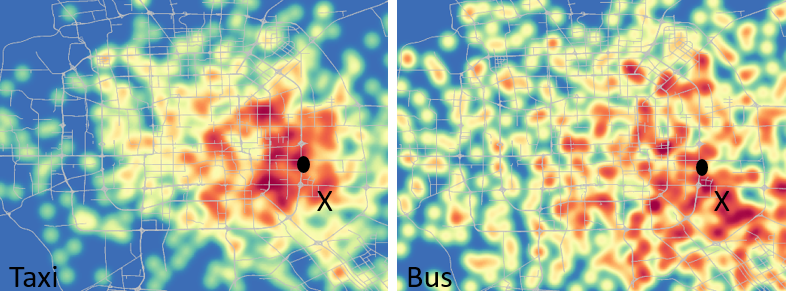}
\caption{The heatmap of the interactions between region X with other regions in the city. A darker colored region indicates a higher probability of people in this region moving to region X.
%The interaction of a region (indicated by ``A'' on the plot) with others.
}~\label{fig:gravity}
\vspace{-2em}
\end{figure}

A multivariate regression method~\cite{burnham1996frameworks} is then employed on (\ref{equ:estimation1}) to estimate the coefficients based on the observed data. We consider two types of commuting in a city: by taxi and by bus. Based them, we compute $q^{B}_{i,j}$ indicating the interaction by bus, and $q^{A}_{i,j}$ indicating the interaction by taxi. 
%We plot these two types of commuting flows to a region in ~\ref{fig:gravity} as an example. 
%Figure~\ref{fig:gravity} shows an example of the quantified interactions from other regions to a region with both taxi and bus data.
Figure~\ref{fig:gravity} shows the interaction between a region (point X on the map) and other regions in the city.
We then fill the quantified interactions into matrices $Q^{B}$ and $Q^{A}$, where the element in the $i$th row and $j$th column is the spatial interaction from region $i$ to region $j$ by bus and taxi, respectively.

\subsubsection{City-wide Interaction Regularization}
Inspired by the social regularization reported in~\cite{ma2011recommender}, we formulate the transitions between regions as a regularization term in our objective functions. As mentioned in the introduction, %across-region transitions carry lifestyle to many regions. 
interactions between regions transit lifestyles too.
The more interactions between two regions, the more alike their lifestyles are. Based on this intuition, we propose the interaction regularization term as:

\begin{equation}
\frac{\alpha}{2} \displaystyle\sum_{i=1}^{r} \Vert R_{l_{i}} -  \frac{1}{2} \displaystyle\sum_{* \in A, B} \frac{\sum_{j=1}^{r}Q^{*}(j,i) \times R_{l_{j}}}{\sum_{j=1}^{r} Q^{*}(j, i)}\Vert_{F}^2,
\end{equation} where $Q^{*}(j, i)$ denotes the interaction from region $j$ to $i$ by bus or taxi, and $R_{l_{i}}$ indicates the lifestyles of $i$th region. 
We take two traffic types equally in the regularization, and average two types of interactions as the final interaction. 
$\alpha$ is a parameter that indicates the importance of this regularization term. This term limits the difference between regions' lifestyles according to their interactions. In other words, a large $Q(j, i)$ leads to a strong influence from region $j$ to region $i$. As a result of introducing this term, the cross-region influence is incorporate to our model.

\subsection{Hybrid Model}
We combine the collective matrix factorization model and interaction regularization term to obtain our final objective function:

\begin{align}\label{equ:loss}
L(R_{l}, V_{1}, V_{2}) \notag\\
= & \frac{1}{2} \Vert  I \circ (R_{s} - R_{l}V_{1}^\top) \Vert_{F}^2 + \frac{\lambda_{1}}{2} \Vert R_{m} - R_{l}V_{2}^\top \Vert_{F}^2\notag\\
&+ \frac{\alpha}{2} \displaystyle\sum_{i=1}^{r} \Vert R_{l_{i}} -   \frac{1}{2} \displaystyle\sum_{* \in A, B} \frac{\sum_{j=1}^{r}Q(j,i) \times R_{l_{j}}}{\sum_{j=1}^{r} Q(j, i)}\Vert_{F}^2\notag\\
&+ \frac{\lambda_{2}}{2}(  \Vert R_{l} \Vert_{F}^2 +  \Vert V_{1} \Vert_{F}^2 +  \Vert V_{2}  \Vert_{F}^2) ,
\end{align} 

There are generally no closed-form solutions for minimizing the above objective function. Therefore, we search for a local minimum by applying gradient descent in the lifestyle vector of each region $R_{l_{i}}$ and matrix $V_{1}$, $V_{2}$ as follows

\begin{align}\label{equ:grad}
 \frac{\partial L}{\partial R_{l_{i}}} = & \displaystyle\sum_{j=1}^{r}I_{i,j}(R_{l_{i}}V_{1_{j}}^\top - R_{s_{i,j}})V_{1_{j}}^\top + \lambda_{2}R_{l_{i}}\notag\\
& + \alpha \displaystyle\sum_{j=1}^{r}(R_{l_{i}} - \frac{1}{2} \displaystyle\sum_{* \in A, B} \frac{\sum_{j=1}^{r}Q^{*}(j,i) \times R_{l_{j}}}{\sum_{j=1}^{r} Q^{*}(j, i)})\notag \\
\frac{\partial L}{\partial V_{1}} =  & [I\circ(R_{l}V_{1}^\top - R_{s})]^\top R_{l} + \lambda_{2} V_{1}\notag\\
\frac{\partial L}{\partial V_{2}} = & \lambda_{1}(R_{l}V_{2}^\top - R_{m})^\top R_{l} + \lambda_{2}V_{2}.
\end{align}

By optimizing the objective function, we obtain $R_{l}$, $V_{1}$, and $V_{2}$. According to Equ.~\ref{equ:approx}, we predict the missing values in $R_{s}$ with the product of $R_{s}$ and $V^{\top}_{1}$. Also, the corresponding rows of $V^{\top}_{1}$ and $V^{\top}_{1}$ explain same lifestyles from two views. We summarize our algorithm in Algorithm~\ref{alg:opt}.

\begin{algorithm}[!h]
%\small
%\SetAlgoLined
\KwIn{region shopping pattern matrix $R_{s}$, region mobility pattern matrix $R_{m}$, region interaction matrix $Q^{A}$ and $Q^{B}$, and parameters: $\lambda_{1}$, $\lambda_{2}$, and $\alpha$}
\KwOut{region lifestyle pattern matrix $R_{l}$, lifestyle-shopping matrix $V_{1}$, and lifestyle-mobility matrix $V_{2}$}
 initialize $R_{l}$, $V_{1}$, and $V_{2}$\;
 //T is maximum iterations, $\epsilon$ is a stopping threshold\\
 \While{$t < T$ and $L_{t} - L_{t+1} > \epsilon $ }{
  calculate the objective function $L_{t}$ based on Equ.~\ref{equ:loss}\;
  calculate $\frac{\partial L}{\partial R_{l_{i}}}$, $\frac{\partial L}{\partial V_{1}}$, and $\frac{\partial L}{\partial V_{2}}$ based on Equ.~\ref{equ:grad}\;
  $\gamma$ = 1\;
  //search for the maximal learning rate\;
  \While{$L(R_{l_{i}}-\gamma\frac{\partial L}{\partial R_{l_{i}}}, V_{1}-\gamma\frac{\partial L}{\partial V_{1}}, V_{2}-\gamma\frac{\partial L}{\partial V_{2}}) \geq L (R_{l}, V_{1}, V_{2})$}{
    $\gamma$ = $\gamma$/2\;
  }
  update $R_{l_{i}}$, $V_{1}$, and $V_{2}$ based on Equ.~\ref{equ:grad}\;
  $t = t + 1$\;
 }
 \Return $R_{l}$, $V_{1}$, and $V_{2}$
 \caption{Optimization of the Hybrid Model}\label{alg:opt}
 %\vspace{-1em}
\end{algorithm}

\section{Evaluation}
\subsection{Data}
We part the city into 1km by 1km grids aligned with the latitude and longitude lines. We take the urban area within Beijing 5th Ring Road, and there are $29 \times 30 = 870$ grids in total, and we do not have any browsing records for 548 (62.9\%) of them. 
Each grid is taken as a region. User online browsing data is used for extracting $R_s$, the shopping patterns of each region. Check-in data from a popular LBSN is used to extract $R_m$, the mobility patterns of regions. To learn the interactions between regions, we utilize two large datasets of taxi and bus transitions in the city. We describe these datasets as follows: 

\begin{itemize}
\item \textit{online browsing dataset}: 
%We collect the online shopping browsing logs of Internet Explorer (IE) users, who have granted permissions for recording their log data. The collecting performed from Nov to Dec 2015, and collected all the log records generated in Beijing. Each record is associated with a product category and a location identifier. There are more than 250 product categories in total. Many constraints are set up in the collection process to protect user privacy, which make this study more challenging. First, there are no user identities in the data. As a result, we are not able to trace any personal browsing history. Also, the locations of records are not precise coordinates, but the locations of network cell towers. This constraint prevents us from locating where the browsing behaviors are taken place exactly.
We collect the online shopping browsing logs of Internet Explorer (IE) users, who have granted permission for recording their log data.  Our dataset is collected from Nov to Dec 2015, and contains all the log records generated in Beijing. Each record is associated with a product category and a location identifier. There are more than 250 product categories in total. Many constraints are set up in the collection process to protect user privacy, which make this study more challenging. First, there are no user identities in the data. As a result, we are not able to trace any personal browsing history. Also, the locations of records are not precise coordinates, but the locations of network cell towers. This constraint prevents us from locating where the browsing behaviors are taking place exactly.

%This dataset contains web browser users' records of visiting online shopping websites. Each record is associated with a product category and a location identifier. There are more than 250 product categories in total. Many constraints are set up in the collection process to protect user privacy, which make this study more challenging. First, there are no user identities in the data. As a result, we are not able to trace any personal browsing history. Also, the locations of records are not precise coordinates, but the locations of network cell towers. This constraint prevents us from locating where the browsing behaviors are taken place exactly.

\item \textit{check-in dataset}: We first collect more than 2 million check-in data in Bejing from Aug 2011 to Jan 2013. A check-in contains a time-stamp, a location, and a user ID. Each location is associated with a POI category, and there are around 200 cateogires in total. Since tourists generate a lot of check-ins in a large city like Beijing, we cannot use these check-ins directly. Following the user IDs, we download the profiles of these users. A profile contains basic information about a user including their home city. We then collect the users, whose home city is Beijing, and aggregate their check-ins. After doing so, we obtain 1.5 million check-ins generated by 43 thousand users.

\item \textit{bus dataset}: This dataset contains 3 million bus-trip records. The data was collected in Beijing from August 2012 to May 2013, and each record is a 5-tuple consisting of a user ID, boarding and alighting locations, and alighting time. 

\item \textit{taxi dataset}: This dataset consists of 1.9 million taxi trips in Beijing from March 2011 to August 2011. Each trip includes the location of the origin and destination, as well as the boarding and alighting time.
\end{itemize}

The check-in data and the transition data are not as up-to-date as our browsing data dataset. Since the urban structure of Beijing has not changed much in recent years, this allows us to apply these data as complementary information sources in our prediction.

\subsection{Metrics}
We apply two popular metrics to evaluate our model, the Mean Absolute Error (MAE) and the Root Mean Square Error (RMSE). RMSE and MEA are defined as:

\begin{align}
RMSE = \sqrt{\frac{1}{T} \displaystyle\sum_{i,j}(R_{s_{i,j}} - \hat{R}_{s_{i,j}})^2},\\ \notag
\\
MEA = \frac{1}{T}\displaystyle\sum_{i,j}|R_{s_{i,j}} - \hat{R}_{s_{i,j}}|,
\end{align} where $R_{s_{i,j}}$ denotes the true value of $j$th shopping pattern in $i$th region, and $\hat{R}_{s_{i,j}}$ is the estimation of it. $T$ is the number of tested values. Clearly, a lower RMSE or MEA value indicates a better performance.

\subsection{Baselines}
We compare our model with three baselines described as follows.
\begin{itemize}

\item \textit{Matrix Factorization (MF)}: this method decomposes $R_{s}$ into $R_{l}$ and $V_{1}$ directly, and uses the product of two results matrices as the prediction.

\item \textit{Collective Matrix Factorization (CMF)}: in this baseline, we factorize $R_{s}$ and $R_{m}$ together to obtain $R_{l}$, $V_{1}$, and $V_{2}$, without using interaction information.

\item \textit{CMF with neighboring information}: since a region usually has the most interactions with the regions surrounded it, in this method, we assume that a region is only affected by neighboring regions. Accordingly, the interaction term is reformed as:
\begin{equation}
\frac{\alpha}{2} \displaystyle\sum_{i=1}^{r} \Vert R_{l_{i}} -  \frac{\sum_{j \in neighbors} R_{l_{j}}}{\Vert neighbors \Vert}\Vert_{F}^2,
\end{equation}where $neighbors$ is the set of indexes of the regions that surround region $i$, and $\Vert neighbors \Vert$ denotes the amount of neighbors. We take the eight surrounding regions as the neighbors of a region. The comparison between our model and this baseline shows the advantage of introducing the interactions between all regions over simply considering local interactions with neighbors. 
\end{itemize}

\subsection{Results}
In many previous works on matrix factorization~\cite{ma2011recommender}, testing data are formed by randomly selected non-zero elements. These selected non-zero elements are compared with the prediction results to evaluate models. In our problem, $R_s$ contains many empty rows. 
%many regions are lack of information, which results in many empty rows. 
To test if our model is able to predict the regions that we do not have any information about, we set up a harsher testing strategy. 
%we manually create some ``empty regions'', and see evaluate our model on them. 
We first randomly select non-zero rows instead of elements from the original matrix to form the testing data. We then compare the non-zero element in these rows with the prediction results. We let $n = 30$, $m = 40$, and $l = 10$, then train the models on 80\% and 90\% of the original data, respectively. We run the experiment ten times, and report the average results in Table~\ref{tab:results}.

\begin{table}
\centering
\begin{tabular}{ | l || c | c || c | c | c | }
\hline

Training Size &  \multicolumn{2}{|c|}{80\%} & \multicolumn{2}{|c|}{90\%}\\
\hline \hline
Metrics & RME & MEA & RME & MEA \\
\hline
MF & 0.235 & 0.443 & 0.213 & 0.420\\
\hline
CMF & 0.230 & 0.408 & 0.203 & 0.400\\
improve& 2.12\%& 7.90\% & 4.69\% & 4.76\%\\
\hline
CMF + N & 0.226 & 0.398 & 0.200 & 0.374\\
improve& 1.73\%& 2.5\%& 1.47\%&6.5\%\\
\hline
CMF + I & 0.223 & 0.394 & 0.197 & 0.367\\
improve& 1.30\% & 1.00\% & 1.50\%&1.87\%\\
\hline
Total& \textbf{5.1}\%& \textbf{10.81}\%& \textbf{7.5}\%&\textbf{12.61}\%\\
\hline

\end{tabular}
\caption{RME and MEA values of each model. ``CMF + N'' denotes CMF model plus neighboring information, and ``CMF + I'' denotes CMF model plus interaction information. ``improve'' denotes the improvement of accuracy of each model compared with the model before it. ``Total'' indicates the improvement of accuracy using our model compared with MF model.
%We also report the boost brought by each model compared with the model before it.
}~\label{tab:results}
\vspace{-2em}
\end{table}

The table shows that our model outperforms all baseline models, and brings a dramatic improvement to the basic matrix factorization model on both metrics. First, introducing mobility patterns into the task with a CMF model brings a large boost to MF. This indicates that the connection between human shopping patterns and mobility patterns works effectively for prediction. Our model learns the connection from the known regions, and leverages it to predict the regions where shopping patterns are unknown. Adding neighboring information to the model also increases performance. Because people prefer to move to a close place~\cite{song2010limits}, a region usually affects the neighboring regions the most. Furthermore, replacing the neighboring information with interactions between regions brings another reasonable boost to accuracy. This proves our assumption that the interactions between regions also spread the shopping patterns of them. In the next section, we report our prediction results, and discuss the geographical distributions of different shopping patterns.

\section{Multiple Views of Lifestyles}

\begin{table*}
\centering
\begin{tabular}{ | l || c | c || c | c || }
\hline

Lifestyle & \multicolumn{2}{c||}{\textbf{Lifestyle 1}} & \multicolumn{2}{c||}{\textbf{Lifestyle 2}} \\
\hline
Views & \textit{Shopping View} & \textit{Mobility View} & \textit{Shopping View} & \textit{Mobility View} \\\hline\hline

%%%%%%%%%%%%%%%%%%%%%%%%%%%%%%%%%%%%%%%%%%%%%%%%%%%%%%%%%%%%%%%%%%%%%%%%%%%%%%%%%%%%%%%%%%%%%%%%%%%%%%%%%%%%%%%%%%%%%%%%%%%%%%%%%%%%%%

\multirow{3}{*}{First Pattern}& Women Clothing & Offices & Tablets & Public Libraries \\
&  Underwear & Convenient Stores & Photography Products & Book Stores  \\
& Women Shoes & Commute & Digital Accessories & Public Art\\ \hline\hline

\multirow{3}{*}{Second Pattern}& Laptops & Apartments \& Communities & Books & Gymnasium \& stadiums  \\
& Office APLS \& Stationery & Commute & Children's Books  & Billiard Parlors\\
& Tablets & Other POIs & Health \& Fitness & School Gyms\\ \hline\hline

\multirow{3}{*}{Third Pattern}& Cellphones & Shopping Malls & Peripheral Products & Digital Appliance Stores  \\
& Cellphone Plans & Cosmetics Stores & Network Products & Commute  \\
& Cellphone Accessories & Sporting Stores & Underwear & Sporting Stores\\ \hline\hline

Lifestyle & \multicolumn{2}{c||}{\textbf{Lifestyle 3}} & \multicolumn{2}{c||}{\textbf{Lifestyle 4}}\\
\hline 
Views & \textit{Shopping View} & \textit{Mobility View} & \textit{Shopping View} & \textit{Mobility View} \\ \hline \hline

%%%%%%%%%%%%%%%%%%%%%%%%%%%%%%%%%%%%%%%%%%%%%%%%%%%%%%%%%%%%%%%%%%%%%%%%%%%%%%%%%%%%%%%%%%%%%%%%%%%%%%%%%%%%%%%%%%%%%%%%%%%%%%%%%%%%%%

\multirow{3}{*}{First Pattern}& Computer Accessories & Furniture Stores & Men Clothing & Cinemas \\  
& Cellphone Accessories & Cosmetics Stores & Men Shoes & Video Game Halls \\  
& Other Computer Products & Garages \& Dealers & Clothing Accessories & Board Games\\ \hline\hline

\multirow{3}{*}{Second Pattern}& Furniture &  Public Squares \& Blocks  & Tablets & Supermarkets \\  
& Home Textiles  & Scenic Spots & Photography Products & Sporting Stores \\ 
& Large Home Appliances & Other POIs & Digital Accessories & Convenient Stores\\ \hline  \hline

\multirow{3}{*}{Third Pattern}& Auto Accessories & Cinema & Books & School Dormitories\\  
& Multimedia products & Video Game Halls & Children's Books & School Laboratories \\
& Cellphone Accessories & Board Games & Health \& Fitness & School Libraries\\   \hline  \hline

\end{tabular}
\caption{Four lifestyles from both shopping and mobility views. For each view we list the top four most weighted shopping/mobility patterns, and for each pattern we list the top four most weighted product/POI categories.``APLS'' is short for ``Appliances''. }~\label{tab:views}
\vspace{-2em}
\end{table*}

In our results, $V_1^{\top}$ and $V_2^{\top}$ indicate shopping and mobility views of latent lifestyles. The reasonable mappings between these views guarantee the effectiveness of incorporating mobility patterns to predict the shopping patterns. In this section, we discuss the mappings discovered in the results. Note that, a shopping view of lifestyle consists of a group of related shopping patterns, and a shopping pattern is formed by a group of related product categories. It is the same for mobility views of lifestyle. Due to the length limit, we pick four latent lifestyles as examples. We list the top three shopping/mobility patterns for the latent lifestyles in Table~\ref{tab:views}. For each shopping/mobility pattern, we list the top three product/POI categories.

%\begin{itemize}
%\item 
\textit{\textbf{Lifestyle 1}} describes the typical life of urban white-collars. The top weighted mobility patterns mainly focus on work (Office, Convenient Stores, and Commute), home (Apartment \& Communities, Commute), and shopping places (Shopping Malls, Cosmetics Stores, and Sporting Stores). Interestingly, instead of office appliances, the most popular shopping pattern is about women's clothing, underwear, and women shoes. This may indicate the online shopping behaviors of office clerks while working. The second shopping pattern consists of Laptops, Office Appliances \& Stationery, and Tablets, which are usually needed for work. The mapping between these two views of this lifestyle explains the overlapping of the two geographical distributions shown in Figure~\ref{fig:compare}. The demands for office appliances and the check-ins at offices can be explained as generated by this latent lifestyle. Therefore, the high level of this lifestyle in a region naturally results in the high levels of both patterns.

%\item 
\textit{\textbf{Lifestyle 2}} shows a strong emphasis on digital and electric products from its shopping view. The top shopping pattern of this lifestyle contains Tablets, Photography, and Digital Accessories, and the third shopping pattern contains Peripheral and Network Products. We can see from its view of mobility that, a group of POI categories lead by Digital Appliance Stores is one of the top mobility patterns. Also, this lifestyle has a high demand for books -- the second shopping pattern is dominated by Books. Correspondingly, we observe that the most popular mobility pattern in this lifestyle contains Public Libraries, Book Stores, and Public Arts. It also shows a preference for gyms (Gym, Stadiums, and Billiard Parlor) of people of this lifestyle. %Although this mobility pattern cannot be derived from the shopping patterns, it is not difficult to imagine that the people having such shopping patterns have a demand for excises. 

%\item 
\textit{\textbf{Lifestyle 3}} is quite life-oriented. We observe the shopping pattern of Furniture, Home Textiles, and Large Home Appliances is listed in the second place in this lifestyle. This pattern implies the need for home arrangements and decorations, and it matches the first mobility pattern of the mobility view, which is lead by Furniture Stores. Also, auto accessories are of high demand in this lifestyle. This corresponds to the visits to garages and dealers in the mobility view. Moreover, the people of this lifestyle have a relatively high preference for traveling, because they pay more visits to scenic spots and public squares (second mobility pattern). This mobility pattern, conversely, implies the demands of auto accessories of the shopping view. The top shopping pattern consists of computer accessories, cellphone accessories, and other computer products, which decries a general demand for electric products.

%\item 
\textit{\textbf{Lifestyle 4}} has the mobility patterns that mainly focus on entertainment (Cinemas, Video Game Halls, and Board Games), shopping places (Supermarkets, Sporting Stores, and Convenient Stores), and schools or universities (School Dormitories, School Laboratories, and School Libraries). From the view of mobility, we suppose that many students could have high rates on this lifestyle. As to its shopping view, Men's Clothing, Male Shoes, and Clothing Accessories form the top shopping pattern of this lifestyle, followed by digital products, and then books. Women's clothing is also quite popular in this lifestyle, but not listed in the top three. For students, it is natural to have high demand for books. This is also reflected in our prediction results -- the level of the shopping pattern of book is high in educational regions (see Section~\ref{sec:book}). Moreover, teenagers usually are more interested in digital products, which explains the digital products shopping pattern. 

%\end{itemize}

The reasonable mapping between two views explains why our model works effectively: a region has certain latent lifestyles, and the views of lifestyles are actually connected. Therefore, we are able to utilize the data of another view (mobility) to prediction our target (shopping).

%In next section, we plot the prediction results of several shopping patterns on a map. Clear differences between their geographical distributions 

\section{Predicted Shopping Patterns in Beijing}

The geographical distributions of shopping patterns not only tell the various demands of regions, but also uncover the urban structure of a city. For example, a high level of shopping behaviors implies a high level of commerce. Furthermore, the level of a specific shopping pattern could reveal key features of regions. In this section, we display our prediction results on a map, and compare the levels of these shopping patterns city-wide. Clear differences between the distributions reveal interesting insights to the city.

\subsection{Business Centers of Beijing}

\begin{figure}[!htbp]
\centering
\includegraphics[width=1 \columnwidth]{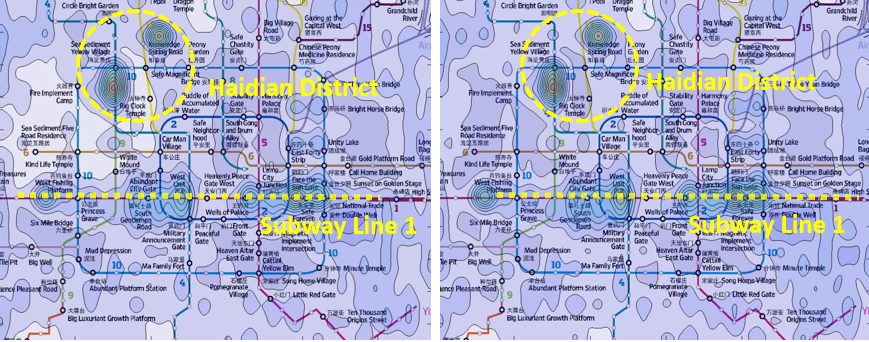}
\caption{The geographical distributions of two shopping patterns on a Beijing subway map. The top three weighted product categories of them are: Laptops, Office Appliances \& Stationery, and Tablets (left); Women Clothing, Underwear, and Women Shoes (right).}~\label{fig:line1}
\vspace{-2em}
\end{figure}

We plot the prediction results of two shopping patterns on a Beijing subway map in Figure~\ref{fig:line1}. The shopping pattern on the left is mainly about office appliances, and the one on the right is mainly about women's clothing. The distributions of these two patterns are quite similar, except that the pattern of women clothing is more widely spread. We suppose that this implies the online shopping behaviors of office clerks during working hours. As we showed in last section, office appliances and women's clothing are clustered into the shopping view of the same latent lifestyle in our results (see Table~\ref{tab:views}, shopping view of lifestyle 1). 

More importantly, the distribution of office-related shopping patterns implies the locations of business centers in the city. 
Several regions have much higher levels on this shopping pattern. Two groups of regions can be observed on the map. The regions in the first group are located in the middle of the city, and they line up horizontally. The line linking these regions aligns with Beijing Subway Line 1, which is highlighted on the figures with a dash line. Line 1 is not only the oldest, but also the most heavily used line of the Beijing Subway~\footnote{https://en.wikipedia.org/wiki/Beijing\_Subway}. It links the earliest and most important economic centers in the city, such as Guomao, Xidan, and Wangfujing. Moreover, the locations of the highest levels of shopping patterns on the line match with the subway stations of Line 1. This concurs with the conclusions on the positive effects of subway stations on the economies of their surroundings~\cite{dewees1976effect}. Also, it can be observed that, the line is split by Tian'anmen Square and the Forbidden City, two landmarks in Beijing.
The levels of this pattern in the east part on this line are higher than the west. This agrees with the fact that the eastern areas are more developed than the western areas in Beijing. According to the statistics from the Beijing Municipal Bureau of Statistics~\footnote{http://old.bjstats.gov.cn/esite/}, this eastern district (Chaoyang district) has the highest Gross Domestic Product (GDP) among all districts in Beijing. The second group consists of the regions in the northwestern corner of the city -- Haidian district (circled on the map). This is a high-tech district hosting many software and computer-technology companies, as well as several higher education institutes~\footnote{https://en.wikipedia.org/wiki/Haidian\_District}. Our results denote that this district has grown to another business center of the city. This agrees with the fact that Haidian takes the second place of the GDP ranking of Beijing. 

By comparing our results with real world locations, we show that our prediction successfully locates the business centers in an urban area.

\subsection{Central VS. Peripheral Regions}

\begin{figure}[!htbp]
\centering
\includegraphics[width=1 \columnwidth]{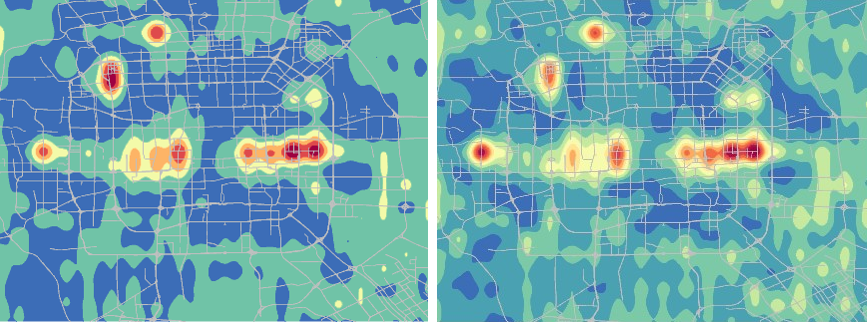}
\caption{The geographical distributions of two shopping patterns. The top three weighted product categories of them are: Furniture, Home Textiles, and Large Home Appliances (left); Baby Products, Toys, and Children Books (right).}~\label{fig:outer}
\vspace{-1em}
\end{figure}

Another two shopping patterns are visualized in Figure~\ref{fig:outer}. The one on the left is about house arrangements and decorations, since Furniture, Home Textiles, and Large Home Appliances are the most weighted product categories of this pattern. The other shopping pattern on the right weights Baby Products, Toys, and Children's Books the most. Not surprisingly, the business centers that we discussed in last section have high levels of these patterns. Because many people work in these areas, the high population base of potential online shopping consumers naturally results in a high level of demand. 

Compared with the shopping patterns of office appliances and women's clothing (See Figure~\ref{fig:line1}), these two patterns are more widely spread. The levels of the pattern of home appliances are relatively high in the peripheral areas of Beijing. As for the patterns of baby products, there are two observations. First, demand for it covers the entire city, which implies that this is a general demand in daily life. Second, the levels in the peripheral regions of this pattern also appear higher than the central areas (except for the business centers). 
The difference between central and peripheral areas reflected by our results can be explained by the process of urban development of Beijing. The central areas of Beijing are traditional districts. These districts are well developed, and the average age of the residents in these areas are relatively high. While the peripheral regions are still under developing. Younger people tend to live in these areas. Also, many ongoing real estate projects are located there. Since products for house arrangements and decorations are usually for new houses, it is reasonable to observe high levels in peripherals. Meanwhile, the baby products are usually purchased by young parents, and the people living in peripheral regions are relatively young, which explains the higher level of the baby product pattern in these regions.

\subsection{Unique Shopping Patterns of Regions}\label{sec:book}
\begin{figure}[!htbp]
\centering
\includegraphics[width=1 \columnwidth]{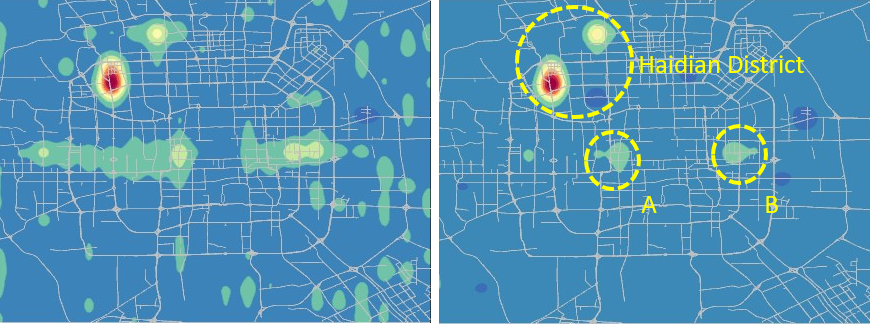}
\caption{The geographical distributions of two shopping patterns. The top three weighted product categories of them are: Desktops, Computer Accessories, and Network Accessories (left); Books, Children's Books, and Health \& Fitness (right).}~\label{fig:haidian}
\vspace{-2em}
\end{figure}

It is not always true that business centers have the highest levels of all shopping patterns. We observe some shopping patterns that are only predicted popular in certain regions. The left plot of Figure~\ref{fig:haidian} describes the geographical distribution of the group for Desktops, Computer Accessories, and Network Accessories. Although this shopping pattern is quite work-related, Haidian district has a much higher level than other business centers. This quite makes sense, because this district is the most developed IT area in Beijing (even in China). The products included in this shopping pattern are especially needed for the IT sector.

Since the need for computers are universal in modern jobs, we can still observe other business centers in the middle of the city in the figure. While another shopping pattern on the right of Figure~\ref{fig:haidian} has the narrowest geographical distribution among all of them.
There are only a few regions have relatively high levels for this shopping pattern. 
%There are only four visible regions on the map. 
Book is the dominant category in this pattern. 
Due to the natural demands for books of students and technicians, Haidian district has the highest demands for books comparing with other regions. Regions A (circled) has the largest book store in Beijing. We suppose this is because people would search books online for extra information such reviews when visiting a bookstore. Region B (cicled) is the Central Business District (CBD) of Beijing. It appears that this region also has a relatively high demands for books.

\section{Conclusion and Future Work}
In this paper, we focus in predicting shopping patterns of different city regions using sparse data. We incorporate two types of extra information to help solve such a challenging problem. First, we connect the shopping patterns with the mobility patterns in a region, in other words, use ``where people go'' to help infer ``what people buy''. Second, we model the interactions between regions, and leverage the information of known regions to infer the shopping patterns in unknown regions. A hybrid model is introduced for the prediction task. The experiment results indicate that incorporating extra information in the task leads to a significant boost (5\%+ in terms of RME, and 10\%+ in terms of MEA) of accuracy for our prediction task. More importantly, our prediction results reveal key functional characteristics of urban areas, including the locations of business centers, the difference between central and peripheral areas, and the unique demands of certain city regions.

There are mainly two directions for our future work. First, we would like to introduce the temporal dimension into the shopping pattern prediction task. Since people's demands vary along time, it is more valuable to predict ``what people buy at when''. Second, we would like to introduce more views of lifestyles, such as personalities, occupations, and so on, to our framework. There are many views to observe how people live. It is therefore beneficial to link more views, and obtain a better and more comprehensive understanding of human lifestyles. 

\section{Acknowledgment}
We would like to thank the Xerox Foundation. We also thank Xitong Yang for helping with the plots in this paper.

\bibliographystyle{abbrv}
\bibliography{sigproc}  

\end{document}